\documentclass{ifacconf}

\usepackage{color}   
\usepackage{graphicx}      
\usepackage{natbib}        

\usepackage{amsfonts}
\usepackage{amsmath} 
\usepackage{amssymb}  

\usepackage{enumitem} 
\usepackage{bbm} 

\newcommand{\vect}[1]{\boldsymbol{#1}}
\newcommand{\mat}[1]{\boldsymbol{#1}}
\newcommand{\wt}[1]{\widetilde{#1}}

\newtheorem{assmp}{Assumption} 
  
\begin{document}
\begin{frontmatter}

\title{On the Analysis of the DeGroot-Friedkin Model with Dynamic Relative Interaction Matrices\thanksref{footnoteinfo}} 

\thanks[footnoteinfo]{This work was supported by the Australian Research Council (ARC) under the ARC grants \mbox{DP-130103610} and \mbox{DP-160104500}, by the National Natural Science Foundation of China (grant 61375072), and by Data61-CSIRO (formerly NICTA). The work of Liu and Ba\c{s}ar was supported in part by AFOSR MURI Grant FA 9550-10-1-0573, and in part by NSF under grant CCF 11-11342.}

\author[First]{Mengbin Ye},  
\author[Second]{Ji Liu}, 
\author[First,Third,Fourth]{Brian D.O. Anderson},
\author[First,Third]{Changbin Yu}, and
\author[Second]{Tamer Ba\c{s}ar}

\address[First]{Research School of Engineering, Australian National University, Canberra, A.C.T. 2601, Australia}
\address[Second]{University of Illinois at Urbana-Champaign, USA}
\address[Third]{Hangzhou Dianzi University, Hangzhou, Zhejiang, China  }
\address[Fourth]{Data61-CSIRO (formerly NICTA Ltd.) }
\address{E-mail: \{mengbin.ye, brian.anderson, brad.yu\}@anu.edu.au, \\
\{jiliu, basar1\}@illinois.edu }

\begin{abstract}                
 This paper analyses the DeGroot-Friedkin model for evolution of the individuals' social powers in a social network when the network topology varies dynamically (described by dynamic relative interaction matrices). The DeGroot-Friedkin model describes how individual social power (self-appraisal, self-weight) evolves as a network of individuals discuss a sequence of issues. We seek to study dynamically changing relative interactions because interactions may change depending on the issue being discussed. In order to explore the problem in detail, two different cases of issue-dependent network topologies are studied. First, if the topology varies between issues in a periodic manner, it is shown that the individuals' self-appraisals admit a periodic solution. Second, if the topology changes arbitrarily, under the assumption that each relative interaction matrix is doubly stochastic and irreducible, the individuals' self-appraisals asymptotically converge to a unique non-trivial equilibrium. 
\end{abstract}


\end{frontmatter}

\section{Introduction}\label{sec:intro}
In the past decade and a half, the systems and control community has conducted extensive research into multi-agent systems. A multi-agent system comprises of multiple interacting agents. Problems such as formation control, distributed optimisation, consensus based coordination and robotics have been heavily studied, see \citep{cao2013,knorn2015_TCNS_MAS_Overview} for two overviews.

On the other hand, the control community has recently turned to study multi-agent systems in social sciences. Specifically, a social network consisting of groups of people interacting with their acquaintances can be considered from one point of view as a multi-agent system. The emergence of social media platforms such as Facebook, Instagram and Twitter has only increased the relevance of research into social networks. 

A problem of particular interest is ``opinion dynamics'', which studies how opinions within a social network may evolve as individuals discuss an issue, e.g. religion or politics. The classical DeGroot model \citep{degroot1974OpinionDynamics} is closely related to the consensus process \citep{jadbabaie2003_CoordinationAgents,shi2013_persistent_social}. Other models include the Friedkin-Johnsen model \citep{FJ}, the Altafini model  \citep{altafini2013antagonistic_interactions,altafini2015predictable_opinions}, and Hegselmann-Krause model \citep{hegselmann2002opinion,etesami}. The DeGroot-Friedkin model proposed and analysed in \citep{jia2015opinion_SIAM} is a two-stage model for \emph{multi-issue} discussions, where issues are discussed sequentially. For each issue, the DeGroot model is used to study the opinion dynamics; each individual updates its own opinion based on a convex combination of its opinion and those of its neighbours. The coefficients of the convex combination are determined by $1)$ the individual's self-weight (which represents self-appraisal, self-confidence) and $2)$ the weights assigned by the individual to its neighbours (which might be a measure of trust or friendship). At the beginning of each new issue, a reflected appraisal mechanism is used by each individual to update its own self-weight. This mechanism takes into account the individual's influence and impact on the discussion of opinions on the prior issue. From one perspective, an individual's self-weight is a representation of that individual's social power in the social network. 

The key contribution of this paper is the study of the DeGroot-Friedkin model with time-varying interactions among the individuals. Because interactions in the DeGroot-Friedkin model are modelled by a matrix termed the ``\emph{relative interaction matrix}'', we will be investigating relative interaction matrices which \emph{dynamically change} between issues but remain constant during each issue. In particular, we will investigate two types of time-varying interactions as to be explained shortly. As an extension of the DeGroot-Friedkin model, a modified version was proposed and analysed in \citep{xu2015_modified_DF, chen2015DF_CT_model}. Time-varying interactions on this modified DeGroot-Friedkin model was studied in  \citep{xia2016_modified_DF_timevary}. On the other hand, there have been no results studying time-varying interactions for the original DeGroot-Friedkin model proposed in \citep{jia2015opinion_SIAM} (which assumed a constant relative interaction matrix during each discussion and between all issues).

The first type of time-varying interaction to consider is issue-dependent interactions that change periodically. For example, consider a government cabinet that meets regularly to discuss several different issues, e.g. defence, finance, and social security. Because different ministers will have different portfolios and specialisations, it is likely that the weights assigned by an individual to its neighbours (used in the opinion dynamics component of the DeGroot-Friedkin model) will change depending on the issue. Periodically changing interactions may occur if a group meets regularly to discuss the same set of issues, e.g. the above example of the government cabinet. Initially, we consider the situation where the social network switches periodically between two different interaction topologies. We show that the self-weight of each individual in the social network has a periodic solution where each individual always has strictly positive self-weight that is less than one. This result is then generalised to multiple periodically switching interaction topologies. 

As a second type of time variation, we consider the case where the issues vary arbitrarily as opposed to periodically. Accordingly, the relative interaction matrices also vary arbitrarily. Furthermore, in order to simplify the analysis, we assume that the arbitrarily varying relative interaction matrices are doubly stochastic \emph{and} irreducible for each issue. Such arbitrarily changing interactions may occur in current day online social networks. We show that, given the assumption of doubly stochastic and irreducible relative interaction matrices, the self-weight of each individual converges asymptotically to $1/n$ where $n$ is the number of individuals in the network. In other words, a democratic sharing of social power is achieved. This result is consistent with \citep{jia2015opinion_SIAM} which studied a single, constant relative interaction matrix for the network over all issues.

The remainder of the paper is organised as follows. Section~\ref{sec:background} provides mathematical notation and introduces the DeGroot-Friedkin model.  Section~\ref{sec:periodic_switch} considers interactions which change periodically with issues. Interactions which vary randomly between issues is studied in Section~\ref{sec:rand_switch}. Simulations are presented in Section~\ref{sec:simulations} and we conclude the paper with the concluding remarks of Section~\ref{sec:conclusion}.

\section{Background and Problem Statement}\label{sec:background}
We begin by introducing some mathematical notations used in the paper. Let $\vect 1_n$ and $\vect 0_n$ denote, respectively, the $n\times 1$ column vectors of all ones and all zeros. For a vector $\vect x\in\mathbb{R}^n$, $0\preceq\vect x$ and $0 \prec \vect x$ indicate component-wise inequalities, i.e., for all $i\in\{1,2,\ldots,n\}$, $0\leq x_i$ and $0<x_i$, respectively. Let $\Delta_n$ denote the $n$-simplex, the set which satisfies $\{\vect x\in \mathbb{R}^n : 0 \preceq \vect x, \vect 1_n^\top \vect x = 1 \}$. The canonical basis of $\mathbb{R}^n$ is given by $\mathbbm{e}_1, \ldots, \mathbbm{e}_n$. Define $\wt{\Delta}_n = \Delta_n \backslash \{ \mathbbm{e}_1, \ldots, \mathbbm{e}_n \}$ and $\text{int}(\Delta_n) = \{\vect x\in \mathbb{R}^n : 0 \prec \vect x, \vect 1_n^\top \vect x = 1 \}$. For the rest of the paper, we shall use the terms ``node'', ``agent'', and ``individual'' interchangeably. We shall also interchangeably use the words ``self-weight'' and ``individual social power''. 

An $n\times n$ matrix is called a {\em row-stochastic matrix} if its entries are all nonnegative and its row sums all equal 1. An $n\times n$ matrix is called a {\em doubly stochastic matrix} if its entries are all nonnegative and its row and column sums all equal 1.

\subsection{Graph Theory}
The interaction between agents in a social network is modelled using a weighted directed graph, denoted as $\mathcal{G} = (\mathcal{V}, \mathcal{E})$. Each individual agent is a node in the finite, nonempty set of nodes $V = \{v_1, \ldots, v_n\}$. The set of ordered edges is $\mathcal{E} \subseteq \mathcal{V}\times \mathcal{V}$. We denote an ordered edge as $e_{ij} = (v_i, v_j) \in \mathcal{E}$, and because the graph is directed, in general the assumption $e_{ij} = e_{ji}$ does not hold. An edge $e_{ij}$ is outgoing with respect to $v_i$ and incoming with respect to $v_j$. The presence of an edge $e_{ij}$ connotes that individual $j$'s opinion is influenced by the opinion of individual $i$ (the precise nature of the influence will be made clear in the sequel). The incoming and outgoing neighbour set of $v_i$ are respectively defined as $\mathcal{N}_i^+ = \{v_j \in \mathcal{V} : e_{ji} \in \mathcal{E}\}$ and $\mathcal{N}_i^- = \{v_j \in \mathcal{V} : e_{ij} \in \mathcal{E}\}$. The relative interaction matrix $\mat C\in\mathbb{R}^{n\times n}$ associated with $\mathcal{G}$ has nonnegative entries $c_{ij}$, termed ``relative interpersonal weights'' in \cite{jia2015opinion_SIAM}. The entries of $\mat C$ have properties such that $0 < c_{ij} \leq 1 \Leftrightarrow e_{ji} \in \mathcal{E}$ and $c_{ij} = 0$ otherwise. It is assumed that $c_{ii} = 0$ (i.e. with no self-loops), and we impose the restriction that $\sum_{j\in\mathcal{N}_i^+} c_{ij} = 1$ (i.e. that $\mat C$ is a row-stochastic matrix).

A directed path is a sequence of edges of the form $(v_{p_1}, v_{p_2}), (v_{p_2}, v_{p_3}), \ldots$ where $v_{p_i} \in \mathcal{V}, e_{ij} \in \mathcal{E}$. Node $i$ is reachable from node $j$ if there exists a directed path from $v_j$ to $v_i$. A graph is said to be strongly connected if every node is reachable from every other node. The relative interaction matrix $\mat C$ is irreducible if and only if the associated graph $\mathcal{G}$ is strongly connected. If $\mat C$ is irreducible then it has a unique left eigenvector $\vect c^\top$ associated with the eigenvalue 1, with the property $\vect c^\top\vect 1_n = 1$ (Perron-Frobenius Theorem, see \citep{godsil2001algebraic}). Hence forth, we shall call this left eigenvector $\vect c^\top$ the \emph{dominant left eigenvector of $\mat C$}.



\subsection{Modelling the Dynamics of the Social Network}\label{ssec:df_model}
The discrete-time DeGroot-Friedkin model is comprised of a consensus model and a mechanism for updating self-appraisal (the precise meaning of self-appraisal will be made clear in the sequel). We define $\mathcal{S} = \{1, 2, 3, \ldots\}$ to be the set of indices of sequential issues which are being discussed by the social network. For a given issue $s$, the social network discusses the issue using the discrete-time DeGroot consensus model. At the end of the discussion (i.e. when the DeGroot model has effectively reached steady state), each individual judges its impact on the discussion (self-appraisal). The individual then updates its own self-weight and discussion begins on the next issue $s+1$. 

\subsubsection{DeGroot Consensus of Opinions}
For each issue $s \in \mathcal{S}$, each agent updates its opinion $y_i(s,\cdot) \in \mathbb{R}$ at the $t+1^{th}$ time instant as
\begin{equation}
y_i(s, t+1) = w_{ii}(s) y_i(s, t) + \sum_{j\in\mathcal{N}_i^+, j\neq i}^n w_{ij}(s) y_j(s, t)
\end{equation}
where $w_{ii}(s)$ is the self-weight individual $i$ places on its own opinion and $w_{ij}$ are the weights given by agent $i$ to the opinions of its neighbour individual $j$. The opinion dynamics for the entire social network may be expressed as 
\begin{equation}
\vect y(s, t+1) = \mat W(s) \vect y(s, t)
\end{equation}
where $\vect y(s, t) = [y_1(s, t) \; \cdots \; y_n(s, t)]^\top$ is the vector of opinions of the $n+1$ agents in the network at time instant $t$. This model was first proposed in \citep{degroot1974OpinionDynamics} with $\mathcal{S} = 1$ (i.e. only one issue was discussed). Below, we provide the model for the updating of $\mat W(s)$ (and specifically $w_{ii}(s)$ via a reflected self-appraisal mechanism). 

\subsubsection{Friedkin's Self-Appraisal Model for Determining Self-Weight} 
The Friedkin component of the model proposes a method for updating the self-weight (self-appraisal, self-confidence or self-esteem) of agent $i$, which is denoted by $x_i(s) = w_{ii}(s) \in [0,1]$ (the $i^{th}$ diagonal term of $\mat W(s)$) \citep{jia2015opinion_SIAM}. Define the vector $\vect x(s) = [x_1(s) \; \cdots \;  x_n (s)]^\top$ as the vector of self-weights for the social network, with starting self-weight $\vect x(1) \in \Delta_n$. The influence matrix $\mat W(s)$ can be expressed as 
\begin{equation}\label{eq:W_matrix}
\mat W(s) = \mat X(s) + (\mat I_n - \mat X(s))\mat C
\end{equation}
where $\mat C$ is the relative interaction matrix associated with the graph $\mathcal{G}$ and $\mat X(s) \doteq diag[\vect x(s)]$. From the fact that $\mat C$ is row-stochastic with zero diagonal entries, \eqref{eq:W_matrix} implies that $\mat W(s)$ is a row-stochastic matrix. The self-weight vector $\vect x(s)$ is updated at the end of issue $s$ as 
\begin{equation}\label{eq:x_update}
\vect x(s+1) = \vect w(s)
\end{equation}
where $\vect w(s)$ is the dominant left eigenvector of $\mat W(s)$ with the properties such that $\vect 1_n^\top \vect w(s) = 1$ and $\vect w(s)\succeq 0 $ \citep{jia2015opinion_SIAM}. This implies that $\vect x(s) \in \Delta_n$ for all $s$. 

In \citep{jia2015opinion_SIAM}, the DeGroot-Friedkin model was studied under the assumption that $\mat C$ was constant for all $t$ and all $s$. In this paper, we investigate the model when $\mat C(s)$ varies between issues. We assume that each agent's opinion, $y_i(s,t)$, is a scalar for simplicity. The results can readily be applied to the scenario where each agent's opinion state is a vector $\vect y_i \in \mathbb{R}^p, p \geq 2$, by using Kronecker products.

It is shown in [Lemma 2.2, \citep{jia2015opinion_SIAM}] that the system \eqref{eq:x_update}, with $\mat C$ independent of $s$, is equivalent to 
\begin{equation}\label{eq:DF_system}
\vect x(s+1) = \vect F(\vect x(s))
\end{equation}
where the nonlinear vector-valued function $\vect F(\vect x(s))$ is defined as
\begin{align}\label{eq:map_F_DF}
\vect F( \vect x(s) ) =   \begin{cases} 
   \mathbbm e_i & \hspace*{-6pt} \text{if } x_i(s) = \mathbbm e_i, \text{for any } i \\ \\
   \alpha (\vect x(s)) \begin{bmatrix} \frac{c_1}{1-x_1(s)} \\ \vdots \\ \frac{c_n}{1-x_n(s)} \end{bmatrix}       & \text{otherwise }
  \end{cases}
\end{align}
with $\alpha(\vect x(s)) = 1/\sum_{i=1}^n \frac{c_i}{1- x_i(s)}$ and where $c_i$ is the $i^{th}$ entry of the dominant left eigenvector $\vect{c}^\top$ of the relative interaction matrix $\mat{C}$. 

\begin{thm}[Theorem 4.1, \citep{jia2015opinion_SIAM}]
For $n\geq 3$, consider the DeGroot-Friedkin dynamical system \eqref{eq:DF_system} with a relative interaction matrix $\mat C$ that is row-stochastic, irreducible, and has zero diagonal entries. Assume that the digraph $\mathcal{G}$ associated with $\mat C$ does not have star topology\footnote{
A graph $\mathcal{G}$ is said to have star topology if there exists a node $i$ such that every edge of $\mathcal{G}$ is either to or from node $i$.} 
and define $\vect c^\top$ as the dominant left eigenvector of $\mat C$. Then,
\begin{enumerate}[label=(\roman*)]
\item \label{prty:thm_DFmain01} For all initial conditions $\vect x(1) \in \wt{\Delta}_n $, the self-weights $\vect x(s)$ converge to $\vect x^*$ as $s\to\infty$. Here, $\vect x^* \in \wt{\Delta}_n $ is the unique fixed point satisfying $\vect x^* = \vect F(\vect x^*)$. 
\item \label{prty:thm_DFmain02} There holds $x^*_i < x^*_j$ if and only if $c_i < c_j$, for any $i,j$, where $c_i$ is the $i^{th}$ entry of the dominant left eigenvector $\vect c$. There holds $x^*_i = x^*_j$ if and only if $c_i = c_j$. 
\item \label{prty:thm_DFmain03} The unique fixed point $\vect x^*$ is determined only by $\vect c^\top$, and is independent of the initial conditions. 
\end{enumerate}
\end{thm}

\subsection{Problem Formulation}
This paper studies the DeGroot-Friedkin model under the assumption that $\mat C$ varies between issues. For a given $s$, however, $\mat C$ is assumed to remain constant for all $t$. Thus, the relative interactions among the individuals, i.e. $\mat C(s)$, may change between issues, but remains constant for all $t$ for a given issue. We will consider alternative situations corresponding to alternative assumptions. We leave the details of these assumptions to their corresponding future sections.

To facilitate our analysis, we make the following two assumptions \emph{which will hold in all problems considered in this paper.}

\begin{assmp}\label{assmp:star}
The graph $\mathcal{G}$ does not have star topology, the relative interaction matrix $\mat C(s)$ is irreducible and $n \geq 3$.
\end{assmp}

\begin{assmp}\label{assmp:initial_conditions}
The initial conditions of the DeGroot-Friedkin model dynamics \eqref{eq:DF_system} satisfy \mbox{$\vect x(1) \in \wt{\Delta}_n$.}
\end{assmp}

Note that Assumption~\ref{assmp:star} requires $n \geq 3$,  because for $n = 2$, any irreducible $\mat C(s)$ is doubly stochastic and corresponds to a star topology. Assumption~\ref{assmp:initial_conditions} ensures that no individual begins with self-weight equal to 1 (autocratic configuration).

\section{Periodic Switching}\label{sec:periodic_switch}
In this section, we investigate the situation where $\mat C(s)$ changes periodically. In order to simplify the problem, we make the following assumption.

\begin{assmp}\label{assmp:2_periodic_top}
The social network switches between two relative interaction matrices, $\mat C_1$ and $\mat C_2$, where both matrices are irreducible, row-stochastic, but not necessarily doubly stochastic. More specifically, the social network switches between $\mat C_1$ and $\mat C_2$ periodically, with period $T = 2$, as given by
\begin{equation}\label{eq:periodic_DF_C}
\mat C(s) = \begin{cases} 
   \mat C_1 &  \text{if } s \text{ is odd} \\
   \mat C_2 &  \text{if } s \text{ is even}
  \end{cases}
\end{equation}
\end{assmp}

Note that for a constant $\mat C$, simulations show that convergence of $\vect x(s)$ to $\vect x^*$ typically occurs after only a few issues \citep{jia2015opinion_SIAM}. We are therefore interested in periodic switching with a short period, because a long period will allow $\vect x(s)$ to reach $\vect x^*$ (hence the above assumption).

\subsection{Transformation into a Time-Invariant System}\label{ssec:transfrom_DF_2top}
Under Assumption~\ref{assmp:2_periodic_top}, the update of the self-weights occurs as $\vect x(s+1) = \vect F(\vect x(s), s)$, where we now acknowledge the fact that $\vect F(\vect x(s), s)$ is an explicit function of time. More specifically, and in accordance with \eqref{eq:map_F_DF}, we have
\begin{equation}\label{eq:periodic_DF_system}
\vect x(s+1) = \begin{cases} 
   \vect F_1(\vect x(s)) &  \text{if } s \text{ is odd} \\
   \vect F_2(\vect x(s)) &  \text{if } s \text{ is even}
  \end{cases}
\end{equation}
The function $\vect F_p$, for $p = 1, 2$, is 
\begin{align}\label{eq:map_Fp_DF}
\vect F_p( \vect x(s)) =   \begin{cases} 
   \mathbbm e_i & \hspace*{-6pt} \text{if } x_i(s) = \mathbbm e_i, \text{for any } i \\ \\
   \alpha_p (\vect x(s)) \begin{bmatrix} \frac{c_{1,p}} {1-x_1(s)} \\ \vdots \\ \frac{c_{n,p}} {1-x_n(s)} \end{bmatrix}       & \text{otherwise }
  \end{cases}
\end{align}
with $\alpha_p(\vect x(s)) = 1/\sum_{i=1}^n \frac{c_{i,p}}{1- x_i(s)}$. Here $c_{i,p}$ is the $i^{th}$ element of the dominant left eigenvector $\vect {c_p}^\top$ associated with the relative interaction matrix $\mat C_p$. 

We now define a new state $\vect y \in \mathbb{R}^{2n}$ as
\begin{equation}\label{eq:y_definition}
\vect y(2s) = 
\begin{bmatrix} \vect y_1(2s) \\ \vect y_2(2s) \end{bmatrix} = \begin{bmatrix} \vect x(2s-1) \\ \vect x(2s) \end{bmatrix}
\end{equation}
and study the evolution of $\vect y (2s)$ for every $s \in \mathcal{S}$. Observe that
\begin{align}\label{eq:y_evolution}
\vect y(2s+2) & = 
\begin{bmatrix} \vect y_1(2s+2) \\ \vect y_2(2s+2) \end{bmatrix} = \begin{bmatrix} \vect x(2s+1) \\ \vect x(2s+2) \end{bmatrix}
\end{align} 
By observing that $\vect x(2s+1) = \vect F_2(\vect x(2s))$ and $\vect x(2s+2) = \vect F_1(\vect x(2s+1))$ for any $s$, we obtain
\begin{equation}
\vect y(2s+2) = 
\begin{bmatrix} \vect F_2(\vect x(2s)) \\ \vect F_1(\vect x(2s+1)) \end{bmatrix}
\end{equation}
Similarly, by noticing that $\vect x(2s) = \vect F_1(\vect x(2s-1))$ and $\vect x(2s+1) = \vect F_2(\vect x(2s))$ for any $s$, we obtain
\begin{align}
\vect y(2s+2) & = 
\begin{bmatrix} \vect F_2\Big(\vect F_1(\vect x(2s-1))\Big) \\ \vect F_1\Big(\vect F_2(\vect x(2s))\Big) \end{bmatrix} \\
& = \begin{bmatrix} \vect F_2\Big(\vect F_1(\vect y_1(2s))\Big) \\ \vect F_1\Big(\vect F_2( \vect y_2(2s) )\Big) \end{bmatrix} \\
& = \begin{bmatrix} \vect F_3 (\vect y_1(2s)) \\ \vect F_4 (\vect y_2(2s)) \end{bmatrix}
\end{align}
for the time-invariant nonlinear composition functions $\vect F_3 = \vect{F}_2 \circ \vect{F}_1$ and $\vect F_4 = \vect{F}_1 \circ \vect{F}_2$. 

We can thus express the periodic system \eqref{eq:periodic_DF_system} as the nonlinear time-invariant system
\begin{equation}\label{eq:TI_DF_system}
\vect y(2s+2) = \bar{\vect F} (\vect y(2s))
\end{equation}
where $\bar{\vect{F}} = [\vect{F}_3^\top, \vect{F}_4^\top]^\top$, and seek to study the equilibrium of this system. More specifically, suppose that $\vect y^* =  [{\vect y_1^*}^\top \; {\vect y_2^*}^\top]^\top$ is an equilibrium of the system \eqref{eq:TI_DF_system}. In Appendix~\ref{app:Lem_F_bar_pf}, we show $\vect{F}_3$ and $\vect{F}_4$ are continuous. It is then straightforward to see that if $\lim_{k\to \infty} \vect y(k) \to \vect y^*$, then $\vect x(s)$ is an asymptotic periodic sequence with periodic sequence  
\begin{equation}\label{eq:periodic_DF_sequence}
\vect x(s) = \begin{cases} 
   \vect F_3(\vect y_1^*) &  \text{if } s \text{ is odd} \\
   \vect F_4(\vect y_2^*) &  \text{if } s \text{ is even}
  \end{cases}
\end{equation}
Define $y_{i}$ (respectively $\bar{F}_{i}$) as the $i^{th}$ element of the vector $\vect y$ (respectively $\bar{\vect F}$). From the above, manipulation allows us to obtain, for $i = 1, \ldots, n$,
\begin{align}
\bar{ F}_{i}(\vect y_1(2s)) & = \alpha_2(\vect F_1(\vect y_1(2s)))
\frac{ c_{i,2} }{ 1 - \alpha_1 (\vect y_1(2s)) \frac{ c_{i,1} } { 1 - y_{i}(2s) } } \label{eq:F_bar_form} 
\end{align}
where $\alpha_1(\vect{y}_1(2s)) = 1/\sum_{j = 1}^n \frac{ c_{j,1} } { 1 - y_{j}(2s) } $ and 
\begin{equation}\label{eq:alpha_2_def}
\alpha_2(\vect F_1(\vect y_1(2s))) = \frac{1}{\sum_{p = 1}^n \frac{ c_{p,2} } { 1 - \alpha_1(\vect{y}_1(2s)) \frac{ c_{p,1} }{1 - y_{p}(2s) } } }
\end{equation}


\subsection{Existence of a Periodic Sequence}

In this subsection, we establish properties of the function $\bar{\vect F}$. More specifically, we detail properties of $\vect F_3(\vect y_1(2s))$. Because $\vect F_3(\vect y_1(2s))$ is similar in form to $\vect F_4(\vect y_2(2s))$, we omit the proof verifying that the same properties hold for $\vect F_4(\vect y_2(2s))$.

\begin{lem}\label{lem:properties_F_bar}
The following properties of $\vect F_3(\vect y_1(2s))$ hold.
\begin{enumerate}[label=P\arabic*]
\item \label{prty:P1} The quantity $\alpha_2(\vect F_1(\vect y_1 (2s) ) ) > 0$ if $\vect y_1(2s) \in \wt{\Delta}_n$, for any $s$.
\item \label{prty:P2} If $\vect y_1(2s) = \mathbbm{e}_i$ for any $i$, then $\vect F_3(\vect y_1(2s)) = \mathbbm{e}_i$. In other words, the $n$ vertices of $\Delta_n$ are fixed points of $\vect F_3$.
\item \label{prty:P3} The function $\vect F_{3}(\vect y_1(2s)) : \Delta_n \rightarrow \Delta_n$ is continuous. 
\item \label{prty:P4} There exists at least one fixed point in $\text{int}(\Delta_n)$.
\end{enumerate}
\begin{pf} 
See Appendix~\ref{app:Lem_F_bar_pf}.
\end{pf}
\end{lem}



Lemma~\ref{lem:properties_F_bar} states that $\vect F_3$ and $\vect F_4$ each have at least one fixed point, which we denote by $\vect y_1^*$ and $\vect y_2^*$ respectively. We will leave the study of whether the fixed points are unique, and analysis of convergence to the fixed points to the journal version of this paper. We now state the following theorem which establishes the periodic behaviour of the system \eqref{eq:periodic_DF_system}. 

%
%

\begin{thm}\label{thm:periodic_sequence} 
Suppose that Assumption~\ref{assmp:2_periodic_top} holds.
\begin{enumerate}[label=T\arabic*]
\item \label{prty:thm_periodic_sequence_01} Suppose further that for some $s_1 \in \mathcal{S}$, there holds $\vect x(2s_1-1) = \vect y_1^*$, where $\vect y_1^* \in \text{int}(\Delta_n)$ is any fixed point of $\vect F_3$. Then, for all $s \geq s_1$, there holds
\begin{equation}\label{eq:periodic_sequence_2}
\vect x(s) = \begin{cases} 
   \vect y_1^* &  \text{if } s \text{ is odd} \\
   \vect y_2^* &  \text{if } s \text{ is even}
  \end{cases}
\end{equation}
where $\vect y_2^* \in \text{int}(\Delta_n)$ is a fixed point of $\vect F_4$.
\item \label{prty:thm_periodic_sequence_02} Suppose now that, instead of \ref{prty:thm_periodic_sequence_01}, there holds for some $s_1 \in \mathcal{S}$, $\vect x(2s_1) = \vect y_2^*$, where $\vect y_2^* \in \text{int}(\Delta_n)$ is any fixed point of $\vect F_3$. Then, \eqref{eq:periodic_sequence_2} holds for all $s_1 \geq s$, with $\vect y_1^* \in \text{int}(\Delta_n)$ being a fixed point of $\vect F_3$. 
\end{enumerate}
\begin{pf} See Appendix~\ref{app:thm_periodic_seq}
\end{pf}
\end{thm}

Note that the above result establishes that a periodic sequence exists, but convergence to this sequence is not established. We conjecture that $\vect F_3$ does in fact have a unique fixed point (i.e. a unique periodic sequence for $\vect{x}(s)$) and that any $\vect y_1(0) \in \wt{\Delta}_n$ will converge to the unique $\vect y_1^*$. We conjecture a similar result for $\vect F_4$. In Section~\ref{sec:simulations}, we provide simulations in support of these conjectures.

\begin{rem}\label{rem:periodic_sequence}
Theorem~\ref{thm:periodic_sequence} leads to an interesting conclusion. Consider that case where, at some point in the evolution of the system trajectory, we have $\vect x(s) = \vect y_1^*$ or $\vect y_2^*$ (e.g. the self-weights are initialised as $\vect x(1) = \vect y_1^*$). Then, the self-weights will exhibit a periodic sequence. Furthermore, for each individual in the network, that individual's self-weight/social power is never zero.
\end{rem}

\begin{rem}\label{rem:nonunique_sequence}
Notice that in the Theorem~\ref{thm:periodic_sequence}, we did not require the fixed points of $\vect F_3$ and $\vect F_4$ to be unique. Suppose that there are two distinct fixed points of $\vect F_3$, which we label $\vect y_{1,a}^*$ and $\vect y_{1,b}^*$. The theorem then concludes that if $\vect x(2s) = \vect y_{1,a}^*$ for some $s$, then the system \eqref{eq:periodic_DF_system} will exhibit a periodic sequence. If on the other hand $\vect x(2s) = \vect y_{1,b}^*$ for some $s$, the system \eqref{eq:periodic_DF_system} will also exhibit a periodic sequence, \emph{but different from the sequence involving $\vect y_{1,a}^*$}.  
\end{rem}

\subsection{Generalisation to $M$ Topologies}\label{ssec:generalise_Mtop}

We now generalise to the case where the social network switches between $M$ different topologies. The following assumption is now placed on the social network instead of Assumption~\ref{assmp:2_periodic_top}.
\begin{assmp}\label{assmp:M_periodic_top}
For $q \in \mathbb{Z}_{\geq 0}$, the social network switches between the $M \geq 3$ relative interaction matrices in the following manner
\begin{equation}\label{eq:periodic_DF_C_M_top}
\mat C(M(s-1) + p) = \mat C_p
\end{equation}
\end{assmp}
for all $s\in \mathcal{S}$ and any $p \in \{1, 2, \ldots, M\}$. The matrices $\mat C_p$ are all irreducible, row-stochastic and in general $\mat C_i \neq \mat C_j,\forall\, i,j \in \{1, 2, \ldots, M\}$.

With the above Assumption~\ref{assmp:M_periodic_top}, the update of the self-weights is given by
\begin{equation}\label{eq:periodic_DF_system_M_top}
\vect x(M(s-1) + p + 1) = 
   \vect F_p(\vect x(M(s-1) + p )) 
\end{equation}
for all $s\in \mathcal{S}$ and any $p \in \{1, 2, \ldots, M\}$. The function $\vect F_p$ is given in \eqref{eq:map_Fp_DF}, but now for $p = 1, 2, \ldots, M$. Following the steps in subsection~\ref{ssec:transfrom_DF_2top}, we now show the generalised transformation of the time-varying system with $M$ different topologies to a time-invariant nonlinear system.

A new state variable $\vect y \in \mathbb{R}^{Mn}$ is defined as 
\begin{equation}\label{eq:y_definition_M}
\vect y(Ms) = 
\begin{bmatrix} \vect y_1(Ms) \\ \vect y_2(Ms) \\ \vdots \\ \vect y_M(Ms) \end{bmatrix} = \begin{bmatrix} \vect x(M(s-1) + 1) \\ \vect x(M (s-1) +2 ) \\ \vdots \\ \vect x(M (s-1) + M) \end{bmatrix}
\end{equation}
and we study the evolution of $\vect y(Ms)$ for every strictly positive integer $s$. It follows that
\begin{align}\label{eq:y_evolution_M}
\vect y(M (s+1)) & = 
\begin{bmatrix} \vect y_1( M(s+1) ) \\ \vect y_2( M(s+1)) \\ \vdots \\ \vect y_M(M(s+1)) \end{bmatrix}
= \begin{bmatrix} \vect x( Ms +1 ) \\ \vect x(Ms+2) \\ \vdots \\ \vect x (Ms + M) \end{bmatrix}
\end{align}
Following the logic in subsection~\ref{ssec:transfrom_DF_2top}, but with the precise steps omitted due to space limitations, we obtain that
\begin{align}
\vect y(M (s+1)) &
 = \begin{bmatrix} \vect F_M ( \vect F_{M-1} (\hdots (\vect F_1 (\vect y_1 ( Ms ) ) ) ) ) \\
\vect F_1 ( \vect F_{M} (\hdots (\vect F_2 (\vect y_2 ( Ms ) ) ) ) ) \\
\vdots \\
\vect F_{M-1} ( \vect F_{M-2} (\hdots (\vect F_M (\vect y_M ( Ms ) ) ) ) ) \end{bmatrix}  \nonumber \\
& \quad = \begin{bmatrix} \vect G_1 (\vect y_1 ( Ms ) ) \\
\vect G_2 ( \vect y_2 ( Ms ) ) \\
\vdots \\
\vect G_M ( \vect y_M ( Ms ) ) \end{bmatrix}
\end{align} 
One can observe that each $\vect G_M$ is a time-invariant nonlinear function. Due to the complexity of each $\vect G_i$, we do not reproduce their expressions here, but their forms are similar to the expressions in \eqref{eq:F_bar_form} - \eqref{eq:alpha_2_def}. The transformed nonlinear system is expressed as
\begin{equation}\label{eq:transformed_DF_M_top}
\vect y(M (s+1) ) = \bar {\vect G}(\vect y(Ms))
\end{equation}

The generalisations of Lemma~\ref{lem:properties_F_bar} and Theorem~\ref{thm:periodic_sequence} are now given below. 
\begin{lem}\label{lem:properties_G_bar}
The following properties of $\bar{\vect G}(\vect y(Ms))$ hold, for any $p\in \{1, 2, \ldots, M\}$.
\begin{enumerate}[label=P\arabic*]
\item \label{prty:P1_G_bar} The quantity $\alpha_j > 0,\forall\, j \in \{1, 2, \ldots, M\}$ if $\vect y_p(Ms) \in \wt{\Delta}_n$, for any $s$.
\item \label{prty:P2_G_bar} If $\vect y_p(Ms) = \mathbbm{e}_i$ for any $i$, then $\vect G_p(\vect y_p(Ms)) = \mathbbm{e}_i$. In other words, the $n$ vertices of $\Delta_n$ are fixed points of $\vect G_p$.
\item \label{prty:P3_G_bar} The function $\vect G_p(\vect y_p(Ms)) : \Delta_n \rightarrow \Delta_n$ is continuous. 
\item \label{prty:P4_G_bar} There exists at least one fixed point for $\vect G_p$ in $\text{int}(\Delta_n)$.
\end{enumerate}
\end{lem}

\begin{thm}\label{thm:periodic_sequence_M}
Suppose that Assumption~\ref{assmp:M_periodic_top} holds. Suppose further that for some $s_1$, there holds $\vect x(M(s_1 -1)+p) = \vect y_p^*$, where $\vect y_p^* \in \text{int}(\Delta_n)$ is a fixed point of $\vect G_p$. Then, for all $s \geq s_1$ there holds
\begin{equation}\label{eq:periodic_sequence_M}
\vect x_j(M(s-1)+j) = \vect y_j^*, \; \text{for any } j\in \{1, 2, \ldots, M\} 
\end{equation}
where $\vect y_j^* \in \text{int}(\Delta_n)$ is a fixed point of $\vect G_j$, and $\vect y_j^* = \vect y_p^*$ for $j = p$.
\end{thm}

The proofs for the above two results are omitted due to the similarity of the proof methods between Lemma~\ref{lem:properties_F_bar} (respectively, Theorem~\ref{thm:periodic_sequence}) and Lemma~\ref{lem:properties_G_bar} (respectively, Theorem~\ref{thm:periodic_sequence_M}).

\section{Arbitrary Switching}\label{sec:rand_switch}

This section considers the scenario where the social network topology changes arbitrarily.
In other words, the function $F(\vect x(s),s)$ in \eqref{eq:map_F_DF} may depend explicitly on the issue $s$. Specifically, it is the quantities $c_i(s)$ that may vary depending on the issue discussed.
In order to facilitate the analysis, the following assumption is invoked.

\begin{assmp}\label{assmp:rand_01}
For all $s\in\{1,2,\ldots\}$, $\mat C(s)$ is doubly stochastic and irreducible.
\end{assmp}




\begin{thm}\label{thm:arb_top_01}
Suppose that Assumption~\ref{assmp:rand_01} holds. Then, the vector of self-weights $\vect x(s)$ asymptotically converges to the unique equilibrium $\vect x^*$ in $\wt{\Delta}_n$ as $s\rightarrow\infty$, where $x^*_i = 1/n$
for each $i\in\{1,2,\ldots,n\}$.

\begin{pf}
Because $\mat C(s)$ is irreducible, the dominant normalised left eigenvector $\vect c(s)^\top$ is unique. Moreover, because $\mat C(s)$ is doubly stochastic, we obtain that $c_i(s) = 1/n$ for all $i\in\{1,2,\ldots,n\}$, 
which implies that $c_i(s)$ does not depend on $s$, and neither does the function $F(\vect x(s),s)$ in \eqref{eq:map_F_DF}.
Thus, the convergence analysis in \citep{jia2015opinion_SIAM}, on the system $\vect x(s+1) = F(\vect x(s))$ can be directly applied here, leading to the conclusion that $\lim_{s\to \infty} \vect x(s) = \vect 1_n /n$.
\end{pf}
\end{thm}

\section{Simulations}\label{sec:simulations}
In this section, simulations are provided which verify the claims of Lemma~\ref{lem:properties_F_bar}, Lemma~\ref{lem:properties_G_bar}, Theorem~\ref{thm:periodic_sequence} and Theorem~\ref{thm:periodic_sequence_M}. The simulated social network has 8 individuals, with three possible sets of interactions described by three different irreducible relative interaction matrices, which we denote as $\mat C_1$, $\mat C_2$ and $\mat C_3$. These are omitted due to space limitations.

\begin{figure*}
\begin{minipage}{0.45\linewidth}
\begin{center}
\includegraphics[width=0.8\linewidth]{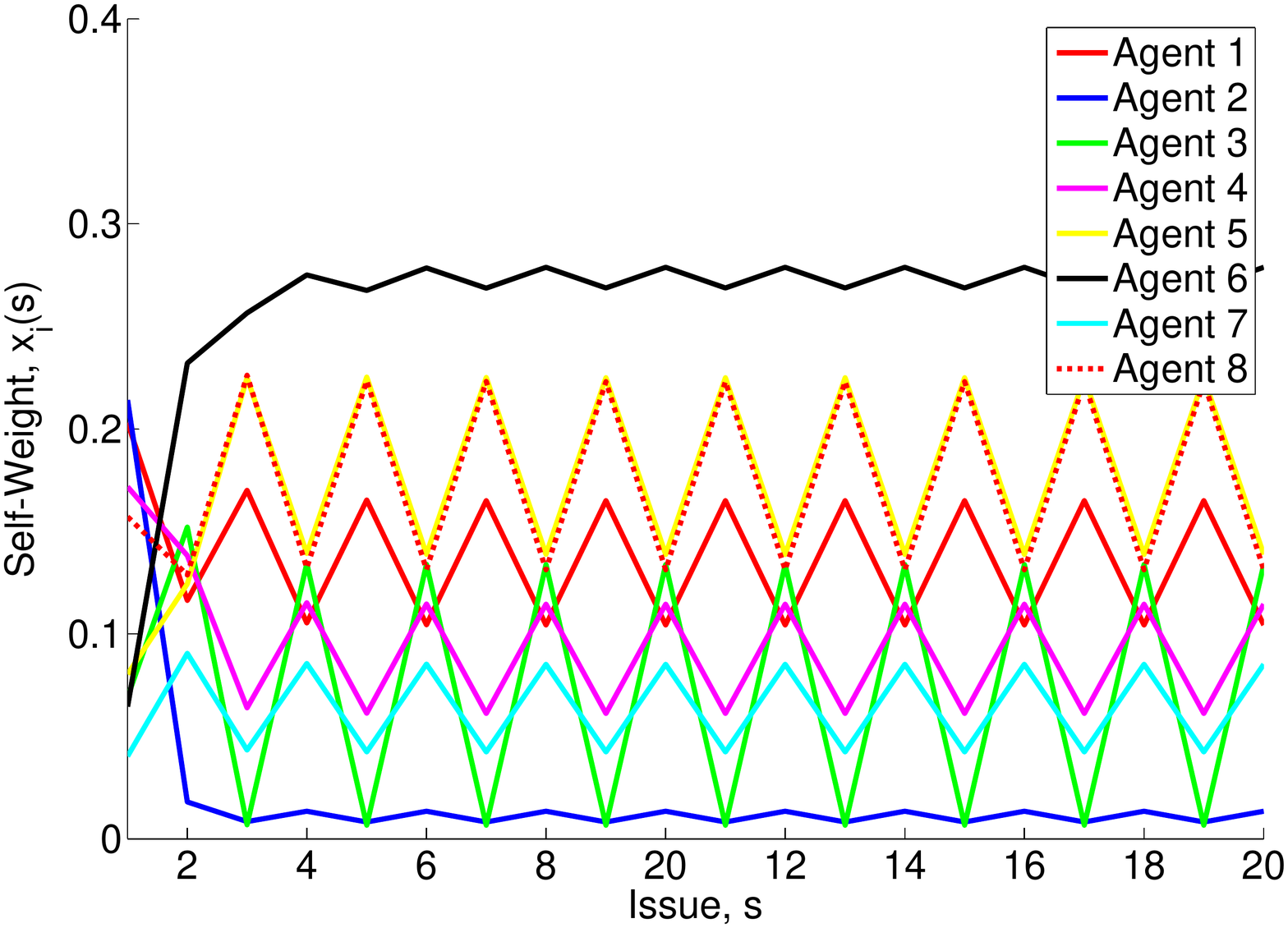}
\caption{Evolution of an individual's self-weight for $\mat C_1$ and $\mat C_2$.}
\label{fig:sim_2top_a}
\end{center}
\end{minipage}
\hfill
\begin{minipage}{0.45\linewidth}
\begin{center}
\includegraphics[width=0.8\linewidth]{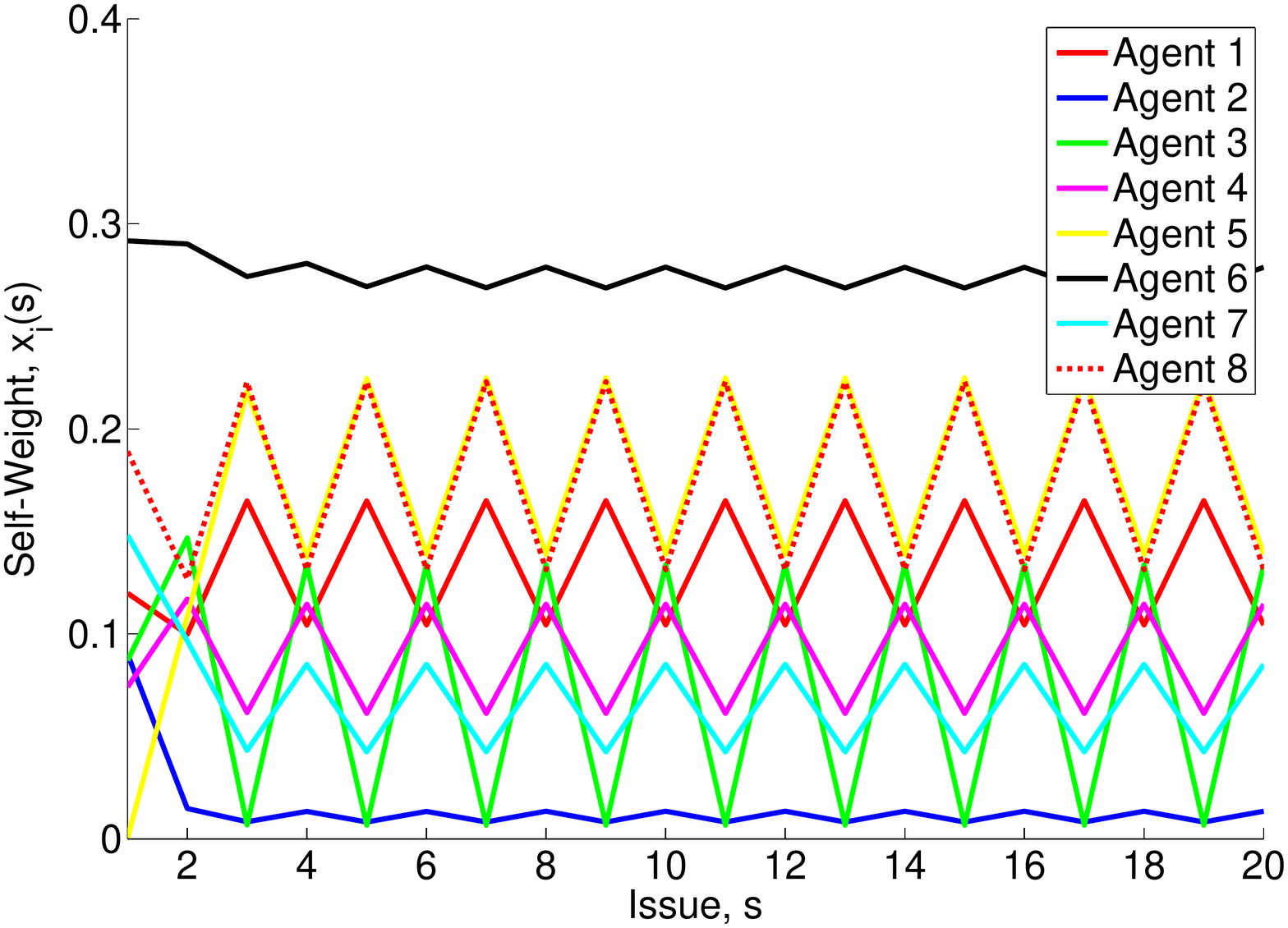}
\caption{Evolution of an individual's self-weight for $\mat C_1$ and $\mat C_2$, different initial conditions.}
\label{fig:sim_2top_b}
\end{center}
\end{minipage}
\end{figure*}

\begin{figure*}
\begin{minipage}{0.45\linewidth}
\begin{center}
\includegraphics[width=0.8\linewidth]{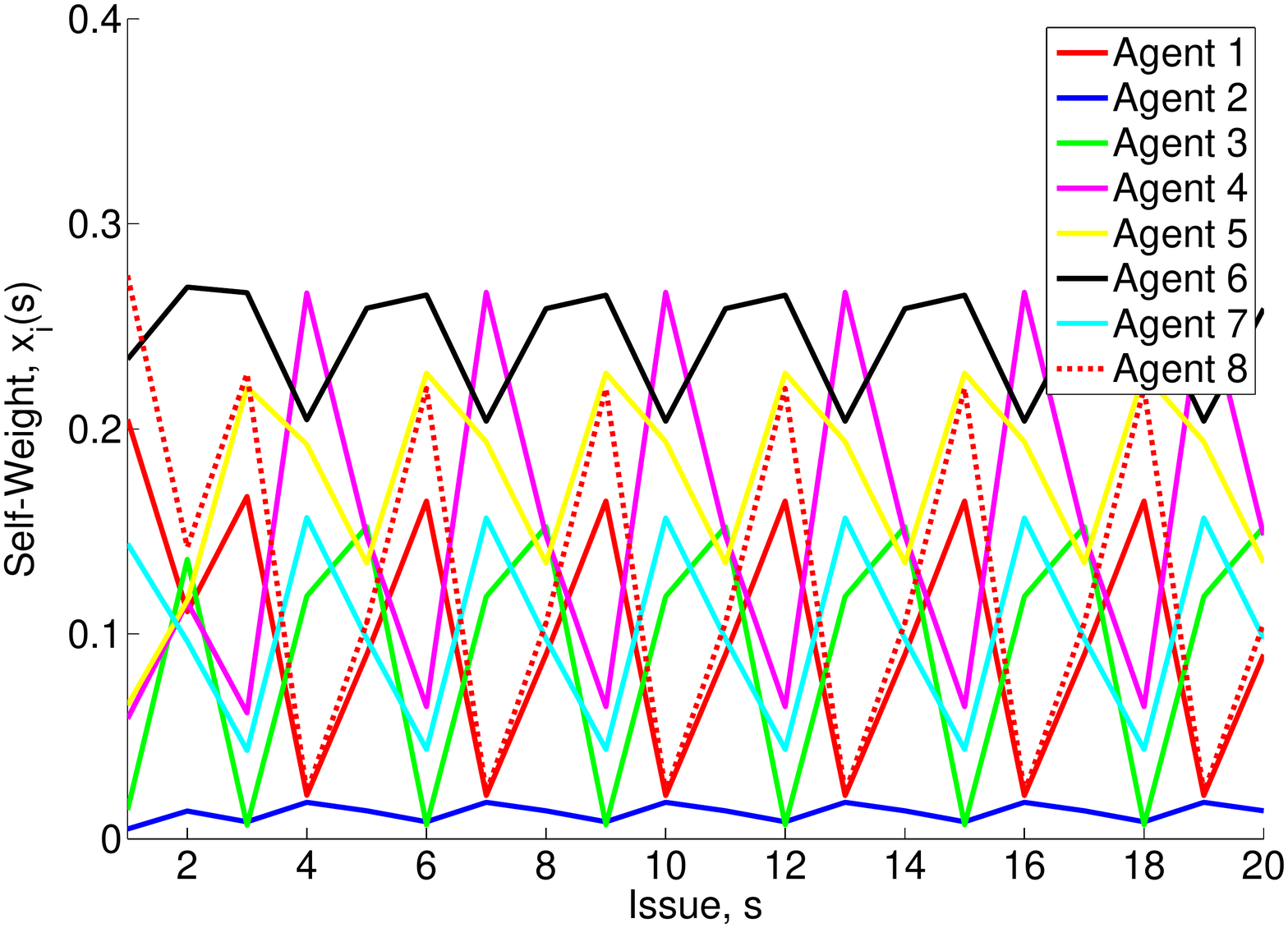}
\caption{Evolution of an individual's self-weight for $\mat C_1$, $\mat C_2$ and $\mat C_3$.}
\label{fig:sim_3top_a}
\end{center}
\end{minipage}
\hfill
\begin{minipage}{0.45\linewidth}
\begin{center}
\includegraphics[width=0.8\linewidth]{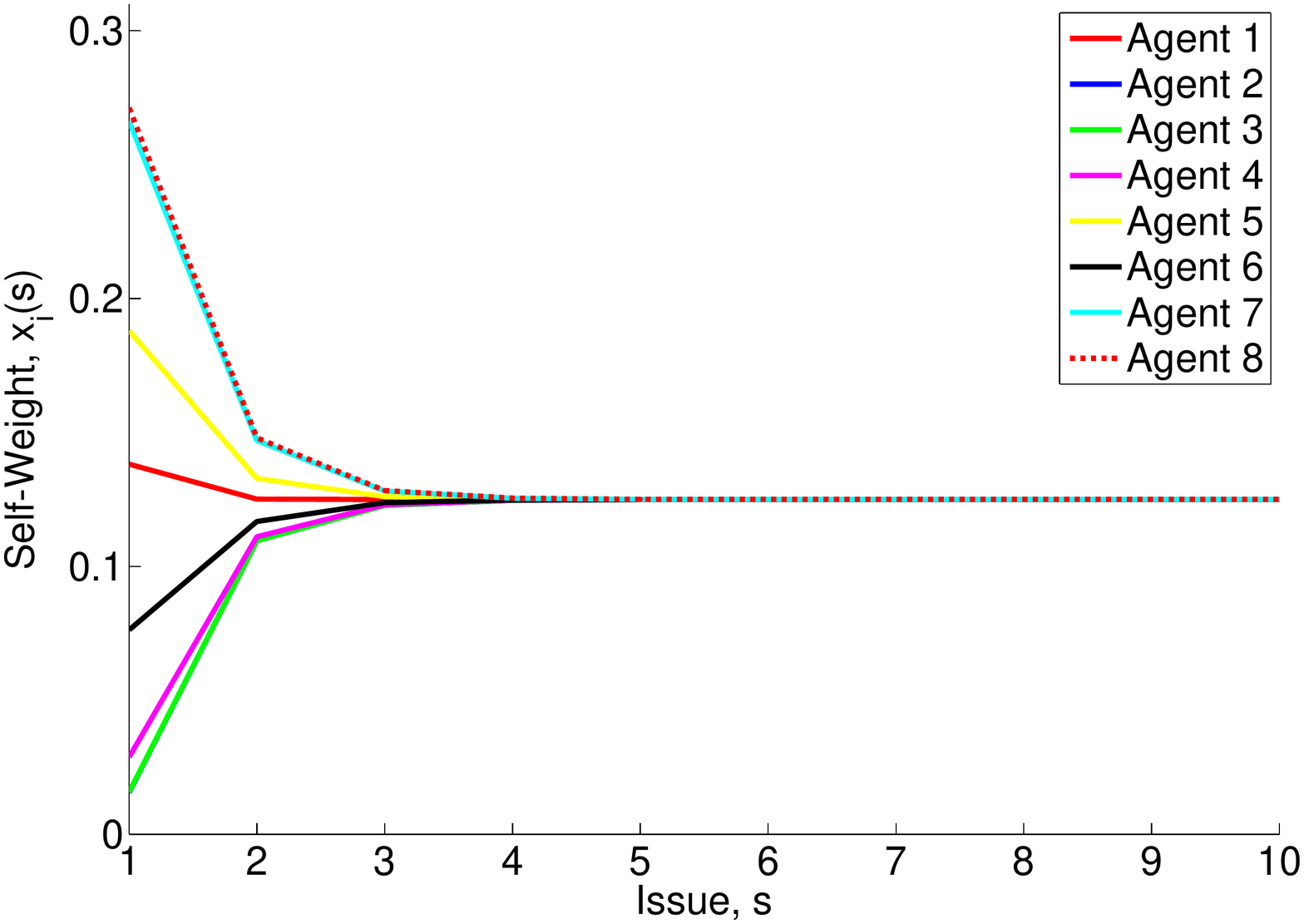}
\caption{Evolution of an individual's self-weight when $\mat C(s)$ varies arbitrarily between $\mat C_4$ and $\mat C_5$.}
\label{fig:sim_arb_top_a}
\end{center}
\end{minipage}
\end{figure*}

Figure~\ref{fig:sim_2top_a} shows the evolution of the individual social power (self-weight $x_i(s)$) over a sequence of issues for the periodically switching relative interaction matrices $\mat C_1$ and $\mat C_2$. The initial condition $\vect x(s = 1)$ was generated randomly. For the same two relative interaction matrices, Fig.~\ref{fig:sim_2top_b} shows the evolution for a second randomly generated initial condition $\vect x(s = 1)$ different from the first figure. Figure~\ref{fig:sim_3top_a} shows the evolution of $\vect x(s)$ for a social network that periodically switches between $\mat C_1$, $\mat C_2$ and $\mat C_3$. 

Figures~\ref{fig:sim_2top_a} and \ref{fig:sim_2top_b} illustrate that Theorem~\ref{thm:periodic_sequence} holds.
In other words, $\vect x(s)$ has a periodic solution. Notice from Figures~\ref{fig:sim_2top_a} and \ref{fig:sim_2top_b} that even for different initial conditions, $\vect x(s)$ asymptotically reaches the same periodic solution. This supports our conjecture that $\bar{\vect F}$ has a unique fixed point and that the fixed point is attractive for all $\vect x(s) \in \text{int}(\Delta_n)$. Our goal is to verify this in the extended version of this paper. Figure~\ref{fig:sim_3top_a} illustrates the results developed in subsection~\ref{ssec:generalise_Mtop} on generalising to multiple periodically switching relative interaction matrices.

Lastly, Figure~\ref{fig:sim_arb_top_a} simulates the DeGroot-Friedkin model for 8 individuals, with $\mat C(s)$ arbitrarily changing between issues. Assumption~\ref{assmp:rand_01} holds for $\mat C(s) \in \{\mat C_4, \mat C_5\}$. Theorem~\ref{thm:arb_top_01} is illustrated since $\lim_{s \to \infty} \vect x(s) \to \vect 1_n/n$.

\section{Conclusion}\label{sec:conclusion}
In this paper, the DeGroot-Friedkin model was used to analyse a social network under a number of assumptions on the dynamically changing network topology, with the network topology being described by relative interaction matrices which vary between issues being discussed by the social network. In particular, results were developed on the evolution of an individual's social power (or self-weight). If the relative interaction matrices changed periodically, then an individual's self-weight admitted at least one periodic solution where the individual's self-weight was always strictly positive, and less than 1. For arbitrarily varying relative interaction matrices which were irreducible and doubly stochastic, individual social power converged to a democratic configuration. In the periodic case, future work will focus on studying the uniqueness of the periodic solution and obtaining a convergence property. For arbitrarily varying topologies, future work will focus on relaxing the assumptions on the relative interaction matrices.


\bibliography{MYE_ANU}             
                                                   







\appendix
\section{Proof of Lemma~\ref{lem:properties_F_bar}}\label{app:Lem_F_bar_pf}

\emph{Property \ref{prty:P1}:} Firstly, observe that 
\begin{equation}\label{eq:alpha_2_01}
\alpha_2( \vect F_1 ( \vect y_1 (2s) ) ) = \frac{1} {\sum_{p=1}^n \frac{c_{p,2} }{1- \bar F_i( \vect y_1(2s) )} }
\end{equation}
and, for $i = 1, \ldots, n$, we have
\begin{equation}
\bar F_i ( \vect y_1(2s) ) = \alpha_1( \vect y_1(2s) )\frac{ c_{i,1} } { 1 - y_i(2s) } 
\end{equation}
From the fact that $c_{p,1} > 0$ and $y_p(2s) < 1$ for any $p = 1, \ldots, n$, we obtain $\alpha_1(\vect y_1( 2s )) > 0$ (see the definition of $\alpha_1$ below \eqref{eq:map_Fp_DF}). It follows that $\bar F_i (\vect y_1 (2s) ) > 0$ for any $i = 1, \ldots, n$. Furthermore, observe that $\sum_{i = 1}^n \bar F_i = 1$ implying that $\vect F_1 (\vect y_1 (2s) ) \in \wt{\Delta}_n$. It is then trivial to obtain from \eqref{eq:alpha_2_01} that $\alpha_2( \vect F_1 ( \vect y_1 (2s) ) ) > 0$, for any $s$.

\emph{Property \ref{prty:P2}:} From \eqref{eq:map_Fp_DF}, it follows that $\vect F_3(\vect y_1(2s)) = \vect F_2( \vect F_1(\mathbbm{e}_i) ) = \vect F_2(\mathbbm{e}_i) = \mathbbm{e}_i$. 

\emph{Property~\ref{prty:P3}:} Let $p = 1,2$. The fact that $\vect F_p : \Delta_n \rightarrow \Delta_n$ is continuous on $\wt{\Delta}_n$ is straightforward; [Lemma 2.2, \citep{jia2015opinion_SIAM}] shows that $\vect F_p$ is Lipschitz continuous about $\mathbbm{e}_i$ with Lipschitz constant $2\sqrt{2}/c_{i,p}$. It is then straightforward to verify that the composition $\vect F_3 = \vect F_2 \circ \vect F_1 : \Delta_n \rightarrow \Delta_n$ is continuous.

\emph{Property~\ref{prty:P4}:} Define the set 
\begin{equation}
\mathcal{A} = \{ \vect y_1 \in \Delta_n : 1 - r \geq   y_i \geq 0, \forall\, i \in \{ 1, \ldots, n \} \}
\end{equation}
where $r$ is strictly positive. In [Theorem 4.1, \citep{jia2015opinion_SIAM}], it is shown that for a sufficiently small $r$, there holds $\vect F_p(\mathcal{A}) \subset \mathcal{A}$, for $p = 1,2$. In fact, $F_i(\vect y_1) < 1 - r,\forall\, i\in \{1, \ldots, n\}$). It follows that $\vect F_1(\mathcal{A}) \subset \mathcal{A} \Rightarrow \vect F_2(\vect F_1(\mathcal{A})) \subset \mathcal{A}$, which implies that $\vect F_3(\mathcal{A}) \subset \mathcal{A}$. 

Brouwer's fixed-point theorem then implies that there exists at least one fixed point $\vect y_1^* \in \mathcal{A}$ such that $\vect y_1^* = \vect F_3(\vect y_1^*)$ because $\vect F_3$ is a continuous function on the compact, convex set $\mathcal{A}$. In the above proof for Property~\ref{prty:P1}, we showed that if $\vect y_1(2s) \in \wt{\Delta}_n$, then $\alpha_1(\vect y_1(2s))$ and $\alpha_2(\vect F_1(\vect y_1(2s)))$ are both strictly positive. For any $i = 1, \ldots, n$, consider now $y_{i}^* = \bar F_i(\vect y_1^*)$. We have
\begin{equation}\label{eq:proof_prty_01}
\bar F_i(\vect y_1^*) = \alpha_2(\vect F_1(\vect y_1^*))
\frac{ c_{i,2} }{ 1 - \alpha_1 (\vect y_1^*) \frac{ c_{i,1} } { 1 - y_{i}^* } }
\end{equation}
which implies that $\bar F_i(\vect y_1^*) > 0$ because $c_{i,p} > 0$ for $p = 1,2$ and $y_i^* < 1$ from the fact that $\vect y_1 \in \wt{\Delta}_n$. In other words, $y_i^*$ is strictly positive for all $i = 1, \ldots, n$. We thus conclude that $\vect y_1^* \in \text{int}(\Delta_n) \subset \mathcal{A}$, i.e. no point on the boundary of $\Delta_n$ is a fixed point apart from $\mathbbm{e}_i$. \hfill  $\square$

\section{Proof of Theorem~\ref{thm:periodic_sequence}} \label{app:thm_periodic_seq}
Observe that $\vect y_1^* = \vect F_2( \vect F_1 (\vect y_1^*) )$. Next, define $\vect z_2^* = \vect F_2(\vect y_1^*)$. We thus have $\vect y_1^* = \vect F_2(\vect z_2^*)$. Next, observe that $\vect F_1 (\vect y_1^*)  = \vect F_1 (\vect F_2 (\vect z_2^*))$ which implies that $\vect z_2^* = \vect F_1 (\vect F_2(\vect z_2^*)) = \vect F_4 (\vect z_2^*)$. In other words, $\vect z_2^*$ is a fixed point of $\vect F_4$. 

For any $s$, suppose that $\vect y_1(2s) = \vect x(2 s - 1) = \vect y_1^*$. From the fact that $\vect x(2s) = \vect F_1(\vect x(2s-1))$, it follows that $\vect y_2(2s) = \vect x(2 s) = \vect z_2^*$. One can then obtain $\vect x(2s + 1) = \vect y_1(2s +2) = \vect F_3(\vect y_1(2s)) = \vect y_1^*$. Likewise, we conclude that $\vect x(2s + 1) = \vect y_2(2s + 2) = \vect F_4(\vect y_2(2s)) = \vect z_2^*$. Because the above arguments hold for any $s$, statement \ref{prty:thm_periodic_sequence_01} in the theorem readily follows. Statement \ref{prty:thm_periodic_sequence_02} follows likewise. \hfill $\square$

\end{document}